\documentclass{aa}

\usepackage{graphicx}
\usepackage{txfonts}
\begin{document} 
   
   \title{Multiwavelength analysis of brightness variations of 3C~279: Probing the relativistic jet structure and its evolution}

\author{Pedro P. B. Beaklini
          \inst{1},
          T\^{a}nia P. Dominici\inst{2}
          \and
          Zulema Abraham\inst{1}
                    \and
          Juliana C. Motter\inst{3}
          }

   \institute{Instituto de Astronomia, Geof\'{\i}sica e Ci\^{e}ncias Atmosf\'{e}ricas,
              Universidade de S\~{a}o Paulo. Rua do Mat\~{a}o 1226, 05508-090, S\~{a}o Paulo/SP, Brazil.\\
              \email{pedro.beaklini@iag.usp.br}
         \and
             Museu de Astronomia e Ci\^{e}ncias Afins, Minist\'{e}rio da Ci\^{e}ncia, Tecnologia, Inova\c{c}\~{o}es e Comunica\c{c}\~{o}es (MAST/MCTIC), Rua General Bruce 586, 20921-030, Bairro Imperial de S\~{a}o Crist\'{o}v\~{a}o, Rio de Janeiro, Brazil
 \and
            Departamento de Astronomia, Universidade Federal do Rio Grande do Sul, Avenida Bento Gon\c{c}alves 9500, 91501-970, Porto Alegre/RS, Brazil\\
             }

   \date{Received xxxxxx xx, xxxx; accepted xxxxx xx, xxxx}
\titlerunning{Brightness variations of 3C~279}
\authorrunning{P.P.B. Beaklini, T. P. Dominici, Z. Abraham, J.C. Motter}

 
  \abstract
   {}
{We studied the correlation between brightness and polarization variations in 3C~279 at different wavelengths,  over time intervals long enough to cover the time lags due to opacity effects. We used these correlations together with VLBI images to constrain the radio and high energy source position.}
   {We made 7 mm radio continuum and $R$-band polarimetric observations of 3C~279 between 2009 and 2014. The radio observations were performed at the Itapetinga Radio Observatory, while the polarimetric data were obtained at Pico dos Dias Observatory, both in Brazil. We compared our observations with the $\gamma$-ray Fermi/LAT  and  $R$-band SMARTS light curves.} 
   {We found a good correlation between 7~mm and $R$-band light curves, with a delay of  $170 \pm 30$ days in radio, but no correlation with the $\gamma$ rays. However, a group of several $\gamma$-ray flares in April 2011 could be associated with the start of the 7 mm strong activity observed at the end of 2011.We also detected an increase in $R$-band polarization degree and rotation of the polarization angle simultaneous with these flares. Contemporaneous VLBI images at the same radio frequency show  two new strong components close to the core, ejected in  directions very different from that of the jet.}
   {The good correlation between radio and $R$-band variability suggests that their origin is synchrotron radiation. The lack of correlation with $\gamma$-rays produced by the Inverse Compton process on some occasions could be due to the lack of low energy photons in the jet direction or to absorption of the high energy photons by the broad line region clouds. The variability of the polarization parameters during flares can be easily explained by the combination of the jet polarization parameters and those of newly formed jet components.}


   \maketitle
%

\section{Introduction}

\label{int}
One of the crucial questions in the understanding of  high energy (HE) emission sources in active galactic nuclei (AGNs) is their actual location. 
Poor spatial resolution at this energy range is compensated by the relatively good time sampling and long-term data series provided by Fermi/LAT\footnote{https://fermi.gsfc.nasa.gov/} since 2008 \citep{atw09}. 
Correlations of HE emission with those at other frequencies also help in the investigation, and radio images obtained with VLBI techniques provide the best spatial resolution.
The radio and millimeter-wave emission clearly originate in a compact, optically thick core and extended jet where bright components move away from the core with apparent superluminal velocities.
The formation of these components has been associated with  the occurrence of optical and infrared flares \citep{mar85};  correlations between optical and $\gamma$-ray events seem to put these emission regions at the same site.
However, their association is not straightforward because opacity effects  introduce time lags between the radio light curves and those  at higher frequencies,  which can compromise  their interpretation and should be analyzed carefully for each source \citep{bot88,ste98,tur99,tur00,cha08,max14,bea14,bea17}. 

3C~279 is the first AGN for which superluminal motions were measured with VLBI techniques \citep{whi71,coh71}, and also the first to be detected as a HE emitter by the Compton Gamma Ray Observatory (CGRO) \citep{har92}.  It is one of the seven flat spectrum radio quasars (FSRQ) detected at very high energy (VHE) with the Cherenkov $\gamma$-ray observatories:  High Energy Stereoscopic System (H.E.S.S.),  Major Atmospheric Gamma Imaging Cherenkov Telescopes (MAGIC), and  Very Energetic Radiation Imaging Telescope Array(VERITAS) \citep{alb08,mar10,mir14,abe15,cer17,mir17}.



3C~279 presents variable emission throughout the entire electromagnetic spectrum on several timescales, showing some periods of low activity and others in which intense flares are observed \citep{mar94,weh98,har01,lar08,cha08,col10,abd10,ale11,hay12,ran17,pat18}. 
It is also a highly polarized blazar, being one of the first radio sources for which optical polarization was measured \citep{kin67}.The polarization is variable on timescales that range from hours \citep{and03} to several months \citep[e.g.,][]{abd10,hay12,hay15,kie16,jer16}, and the polarization degree ($PD$)  varies from values smaller than 1\% to values as high as 45\%. 
  
 A historical optical light curve of 3C~279, starting in 1927, was compiled by \citet{web90}, who noticed a similarity in the flaring activity at two epochs separated by 50 years; \citet{fan99}, using 27 years of observations in the near infrared, found evidence of a 7-year periodicity in the light curve. Those periodicities, together with variations in the velocities and position angles of the superluminal jet components, were interpreted in terms of  jet precession models \citep{abr98,qui11}.

3C~279 was included in the blazar sequence according to the classification scheme of \citet{GHI98}, based on the HE properties of its spectral energy distribution (SED). Its two-peaked SED is a common feature in the blazar class. The first peak, between radio and X-rays, is attributed to synchrotron radiation of relativistic electrons, while the second one at $\gamma$-ray energies, is probably due to the inverse Compton process involving high energy electrons and low energy photons, either from the synchrotron emission (synchrotron self-Compton, SSC) or from an external source (external Compton, EC), although hadronic models are sometimes necessary to explain the VHE emission  \citep[e.g.,][]{man93,bot09,bot13,liu19}.
When trying to understand the SED of blazars, it is necessary to separate the contribution of their quiescent or slowly varying source from the flaring components.
In 3C~279, the quiescent SED was identified from its lowest emission states (January 1993 and January 1995),  which even allowed the study of the accretion disk, as reported by \citet{pia99}.



Since the launch of the Fermi Space Observatory the efforts to observe the counterparts of the $\gamma$-ray emission in 3C~279 at lower frequencies have been increasing \citep[e.g.,][]{lar08,cha08,col10,hay12,ran17}.
Multiwavelength observations have shown evidence of correlations between $\gamma$-ray and optical flares, with delays from 1 to 10 days between them \citep{har01b,abd10,hay12,ran17}. 
This kind of correlation is not restricted  to the optical flux density alone, but also involves  variations in the optical  {\it PD} and polarization angle ({\it PA}) \citep{abd10}.
However, the relation between $\gamma$-ray flares, radio and millimeter  wave emission is still not clear, probably due to   large time lags between the events at low and high frequencies, and the possible superposition of emission of different flares at the lower frequencies.

In this paper, we report the radio and optical polarimetric variability of 3C~279 between 2009 and  2014, with emphasis on the very intense activity at radio frequencies that started at the end of 2011 and lasted for almost a year that did not seem to be correlated with any $\gamma$-ray activity. In Section \ref{obs} we discuss the observational methods used in the radio observations with the Itapetinga radiotelescope and the polarimetric observations at the Pico dos Dias Observatory (OPD). In Section \ref{res} we present our observational results, and in Section \ref{dis} our interpretation. Finally, in Section \ref{conclusion} we  state our conclusions. 
   

\begin{center}

\begin{figure*}
  \includegraphics[width=18cm]{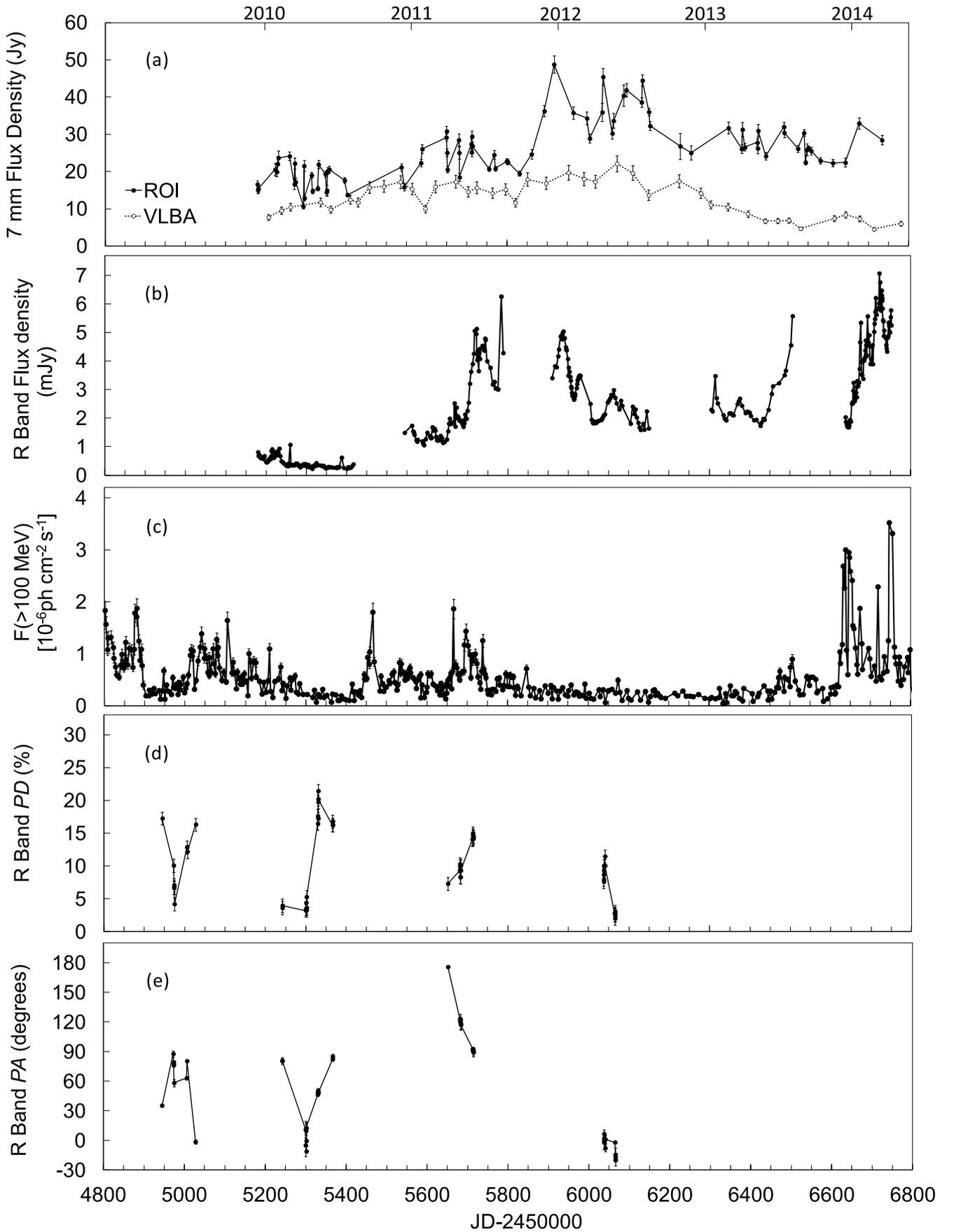}
  \caption{ Variability in 3C279 between 2009 and 2014. (a): 7mm single-dish light curve obtained in this work together with flux per beam of the peak obtained using VLBA data from the VLBA-BU BLAZAR monitoring project; (b): $R$-band photometry from SMARTS; (c): Fermi/LAT $\gamma$-ray light curve from \cite{hay15}; (d): $R$-band polarization degree ($PD$) and position angle ($PA$) obtained in this work.}
  \label{all}
\end{figure*}

\end{center}


\section{Observations}
\label{obs}


The original 7 mm data presented in this work were obtained from the blazar monitoring program performed between 2009 and 2014  (MJD 55179 to MJD 56733) at the Itapetinga Radio Observatory\footnote{Operated by  Instituto Nacional de Pesquisas Espaciais (INPE/MCTIC)} (ROI) and the $R$-band polarimetry  at the Pico dos Dias Observatory\footnote{ Operated by the Laborat\'orio Nacional de Astrof\'\i sica (LNA/MCTIC) }, both in Brazil. Data at the two observatories were obtained simultaneously when possible, on a monthly basis. Previous results of this program were published by \citet{bea14} and \citet{bea17}, corresponding to data analysis of the sources 3C~273 and PKS1510-089, respectively. The observational methods were already  described in detail in those papers and will be summarized here. 

\subsection{7 mm observations at the Itapetinga Radio Observatory}
\label{radio_obs}

The 7 mm  observations were made with the 14 m radome enclosed radiotelescope, which gave a HPBW of 2.4 arc min. The on-the-fly observing method was used, in which 30 scans centered at the source were made, each  with 30 arc min amplitude and 20 s duration. The scan direction was switched between  azimuth and elevation to check the pointing accuracy. On a typical day between 6 and 14 observations were obtained for each scan direction. A room temperature, 1 GHz double side band K-band receiver was used, with  noise temperature of about 700~K. A room temperature load and a noise source of known temperature were used for instrumental calibration and correction of atmospheric absorption in the presence of the radome \citep{abr92}. The instrumental calibration was repeated every 30 minutes. Absolute flux calibration was carried out daily, using the  galactic source SgrB2 Main. The full dataset at 7 mm is presented in Table 1.

\subsection{Optical polarimetry at \textit{Pico dos Dias} Observatory} 
\label{pol}

Optical  polarimetric observations of 3C~279 were carried out between 2009 and 2012  (MJD 54944 to MJD 56065), using the 0.6 m  Boller \& Chivens IAG/USP telescope  and the imaging polarimeter IAGPOL \citep{mag96} working in linear polarization mode and with a standard $R$-band filter. The polarimeter consists of a rotatable, achromatic half-wave retarder, followed by a calcite Savart plate,  such  that it provides two images of each object in the field, with orthogonal polarizations between them, separated by $25.5$ arcsec at the telescope focal plane. The simultaneous detection of the two beams allows observations under non-photometric conditions, and has the advantage that the sky polarization is practically canceled out. On photometric nights, the total flux density can be recovered by adding the two polarimetric components of a given image \citep{bea17}. 


Two different CCDs were used throughout the monitoring program: a $1024 \times 1024$ pixel CCD of $24$ microns per pixel and a $2048 \times 2048$ pixel CCD of $13.5$ microns per pixel, both providing a field of view of about $10^{'} \times 10^{'}$ ($0$\farcs $67$/pixel and $0$\farcs $38$/pixel, respectively). On one typical night, between $1$ and $5$ polarimetric measurements were obtained, each  consisting of eight images obtained from different wave plate positions separated by $22^{\rm o}.5$, consuming a mean total integration time of about 30 minutes, depending on sky quality.

The images were reduced with IRAF\footnote{IRAF is distributed by the National Optical Astronomy Observatory, which  is  operated  by  the  Association  of  Universities  for  Research in  Astronomy,  Inc., under  cooperative  agreement  with  the  National Science Foundation} usual routines to apply bias and flat-field corrections, while the polarimetric data were obtained using the PCCDPACK package \citep{per00}. The polarized standard stars used were HD$298383$, HD$111579,$ and HD$155197$ \citep{tur90}, while the lack of instrumental polarization was confirmed by observing the unpolarized stars HD$94851$ \citep{tur90} and WD$1620-391$\footnote{From the IAG/USP polarimetric group, available at: http://www.astro.iag.usp.br/~polarimetria/padroes/index.html}. To compute calibrated flux densities throughout this work, a galactic extinction of $A_{R} = 0.076$ was considered \citep{sch98}. The full dataset of our polarimetric measurements is presented in Table \ref{151}.\footnote{In Table \ref{151} we present the $PA$ values between $0^{\rm o}$ and $180^{\rm o}$. In Fig. \ref{all} we allowed values higher than $180^{\rm o}$ to avoid artificial jumps in $PA$.}

\begin{figure}
  \includegraphics[width=\columnwidth]{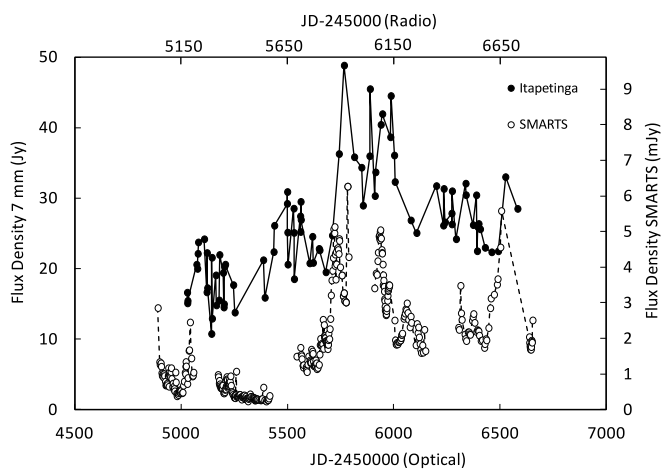}
  \caption{Light curves of 3C~279 at radio (7 mm) and optical ($R$-band) wavelengths. The optical light curve scale is shifted by 150 days relative to radio. }
  \label{7R}
\end{figure}

\begin{figure}
\center 
  \includegraphics[width=8.5 cm]{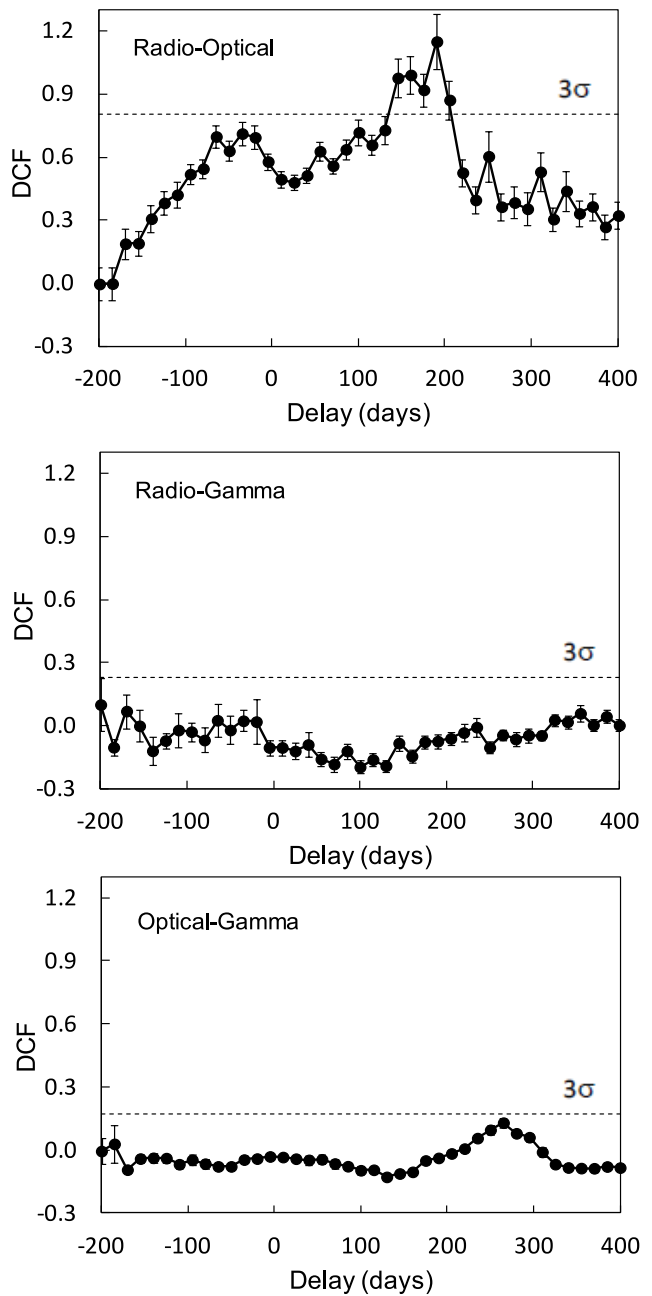}
  \caption{ DCF result between radio and optical (top), radio and $\gamma$ rays (middle), $\gamma$ rays and optical (bottom). The dashed line indicates the 3$\sigma$ level. A significant peak at $170 \pm 30$ day can be seen in the DCF between the radio and optical light curves.}
  \label{DCF}
\end{figure}




\section{Results}
\label{res}

\begin{table*}
 {\small
 \caption{Radio flux density obtained in this work.}
 \hfill{} 
 \begin{tabular}{ c c c c c c c c}
 \hline Day & JD-2400000 & Flux Density (Jy) & Error & Day & JD-2400000 & Flux Density (Jy) & Error \\
\hline 2009-12-14 & 55179 & 16.59 & 0.99 & 2011-08-26 & 55799 & 22.81 & 0.78    \\
2009-12-15 & 55180 & 15.07 & 0.92 & 2011-08-29 & 55802 & 22.60 & 0.72   \\
2009-12-16 & 55181 & 15.42 & 1.01 & 2011-09-27 & 55831 & 19.48 & 0.71   \\
2010-01-27 & 55223 & 20.56 & 1.64 & 2011-10-27 & 55861 & 24.69 & 1.24   \\
2010-01-31 & 55227 & 19.97 & 1.27 & 2011-11-28 & 55893 & 36.26 & 1.55   \\
2010-02-02 & 55229 & 22.11 & 1.97 & 2011-12-22 & 55917 & 48.82 & 2.27   \\
2010-02-04 & 55231 & 23.72 & 1.92 & 2012-02-08 & 55965 & 35.81 & 1.68   \\
2010-03-04 & 55259 & 24.17 & 1.11 & 2012-03-13 & 55999 & 34.33 & 1.77   \\
2010-03-16 & 55271 & 16.64 & 1.41 & 2012-03-20 & 56006 & 28.93 & 1.22   \\
2010-03-17 & 55272 & 22.22 & 1.42 & 2012-04-20 & 56037 & 35.96 & 2.42   \\
2010-03-19 & 55274 & 17.23 & 1.03 & 2012-04-22 & 56039 & 45.48 & 2.29   \\
2010-04-07 & 55293 & 10.73 & 0.77 & 2012-05-15 & 56062 & 30.30 & 1.58   \\
2010-04-09 & 55295 & 21.55 & 1.43 & 2012-05-18 & 56065 & 33.67 & 1.97   \\
2010-04-10 & 55296 & 12.89 & 0.67 & 2012-06-12 & 56090 & 40.41 & 2.85   \\
2010-04-28 & 55314 & 19.05 & 0.79 & 2012-06-19 & 56097 & 41.93 & 1.77   \\
2010-04-30 & 55316 & 14.78 & 0.64 & 2012-07-27 & 56135 & 38.64 & 1.27   \\
2010-05-13 & 55329 & 15.49 & 0.69 & 2012-07-29 & 56137 & 44.50 & 1.51   \\
2010-05-15 & 55331 & 21.94 & 1.12 & 2012-08-14 & 56153 & 36.05 & 1.00   \\
2010-06-01 & 55348 & 19.39 & 0.89 & 2012-08-18 & 56156 & 32.30 & 1.19   \\
2010-06-04 & 55351 & 14.48 & 0.94 & 2012-10-31 & 56231 & 26.84 & 3.52   \\
2010-06-05 & 55352 & 14.96 & 1.45 & 2012-11-27 & 56258 & 25.05 & 1.94   \\
2010-06-07 & 55354 & 20.25 & 0.92 & 2013-02-28 & 56351 & 31.72 & 1.59   \\
2010-06-11 & 55358 & 20.57 & 0.94 & 2013-04-02 & 56384 & 26.10 & 1.11   \\
2010-07-19 & 55396 & 17.66 & 0.79 & 2013-04-04 & 56386 & 31.31 & 1.90   \\
2010-07-26 & 55403 & 13.76 & 0.44 & 2013-04-10 & 56392 & 26.56 & 1.03   \\
2010-12-07 & 55537 & 21.20 & 0.98 & 2013-05-11 & 56423 & 27.82 & 0.93   \\
2010-12-14 & 55544 & 15.86 & 1.05 & 2013-05-12 & 56424 & 26.27 & 1.45   \\
2011-01-25 & 55586 & 22.35 & 0.98 & 2013-05-13 & 56425 & 30.98 & 1.69   \\
2011-01-28 & 55589 & 26.09 & 1.18 & 2013-06-01 & 56444 & 24.17 & 0.96   \\
2011-03-29 & 55649 & 29.20 & 0.95 & 2013-07-16 & 56489 & 32.06 & 1.19   \\
2011-03-30 & 55650 & 30.87 & 1.39 & 2013-07-17 & 56490 & 30.43 & 1.27   \\
2011-03-31 & 55651 & 25.11 & 1.01 & 2013-08-20 & 56524 & 26.18 & 0.92   \\
2011-04-01 & 55652 & 20.57 & 0.92 & 2013-09-04 & 56539 & 30.41 & 0.89   \\
2011-04-29 & 55680 & 28.51 & 1.64 & 2013-09-08 & 56543 & 22.47 & 0.61   \\
2011-04-30 & 55681 & 25.07 & 1.63 & 2013-09-11 & 56546 & 25.87 & 1.63   \\
2011-05-01 & 55682 & 18.50 & 1.12 & 2013-09-15 & 56550 & 26.34 & 0.68   \\
2011-05-30 & 55711 & 27.42 & 1.31 & 2013-09-22 & 56557 & 25.58 & 1.07   \\
2011-05-31 & 55712 & 25.20 & 1.2 & 2013-10-15 & 56580 & 22.96 & 0.83    \\
2011-06-01 & 55713 & 29.48 & 1.51 & 2013-11-15 & 56611 & 22.32 & 0.91   \\
2011-06-02 & 55714 & 26.91 & 1.13 & 2013-12-16 & 56642 & 22.46 & 1.20   \\
2011-07-13 & 55755 & 20.75 & 0.68 & 2014-01-19 & 56676 & 32.99 & 1.50   \\
2011-07-26 & 55768 & 24.53 & 1.18 & 2014-03-17 & 56733 & 28.48 & 1.27   \\
2011-07-29 & 55771 & 20.81 & 0.65 &  &  &  &    \\

\hline   
\hline
\end{tabular}}
\hfill{}
\label{radio_table}
\end{table*}


In Fig. \ref{all} we present our observations together with data at other wavelengths obtained from the literature. 
In the top panel (a) we show the 7 mm light curve obtained with the Itapetinga radiotelescope between 2009 and 2014  and the contemporaneous 7 mm flux density of the VLBI core, obtained from the VLBA-BU-BLAZAR Program\footnote{http://www.bu.edu/blazars/VLBAproject.html}. A detailed description of that program can be found at \citet{jor16}. 

In the same Figure (b) we  show the $R$-band light curve obtained from SMARTS\footnote{http://www.astro.yale.edu/smarts/} Optical/IR Observations of Fermi Blazars\footnote{http://www.astro.yale.edu/smarts/glast/home.php}. That  program operates two small aperture telescopes located at Cerro Tololo Observatory to perform the photometric monitoring of a sample of blazars at $B$, $V$, $R$, $J$, and $K$ bands \citep{smarts}. 

The light curve showing the source activity at $\gamma$ rays in the same time span can be seen in the middle panel of Fig.\ref{all}  (c). To build the light curve, the Fermi/LAT data between 0.1 and 300 GeV were binned in intervals of 3 days in order to increase the S/N ratio \citep{hay15}. Finally, the variability in optical polarization degree  and position angle obtained from our observations are shown in the two bottom panels  (d and e). 

Through the analysis of our single-dish 7 mm light curve we can see a systematic increase in the flux density along the years, and the superposition of  several flares on shorter timescales. The strongest flare  was observed in December 2011  (MJD 55917), when 3C~279 reached a maximum flux density of $48.8 \pm 2.3$ Jy.The minimum flux density of $10.7 \pm 0.8$ Jy was observed in 7 April 2010  (MJD 55293). The core emission obtained from VLBA images at the same frequency was generally smaller than that obtained in the single-dish observations, showing that  a large part of the emission was produced in the parsec-scale jet, especially during the flares observed in 2011-2012.   

At $\gamma$-ray wavelengths we see a large number of short timescale fluctuations  at the beginning of the Fermi/LAT observations, and an isolated flare in 25 September 2010 (MJD 55465),  followed by a small group of fast flares beginning in 15 April 2011 (MJD 55667), and lasting for approximately 70 days. This last group of flares occurred about 5 months before the beginning of the strong radio flare, reaching a flux of $(1.8 \pm 0.3) \times 10^{-6}$ ph s$^{-1}$ cm$^{-2}$. At that moment, it was the highest flux detected at $\gamma$-ray wavelengths, but it was still an order of magnitude weaker  than the flares detected several years later. A period of very low activity followed, which lasted until  2014 when  a large number of high intensity $\gamma$-ray flares was detected \citep{hay15,pal15,pal16,ran17,pat18}; it was, unfortunately, at the end of our monitoring. Differently from the $7$mm single-dish light curve, we do not see any systematic increase in the $\gamma$-ray emission along the years.  

At the $R$ band,  high activity occurred between 2011 and 2012, with several short-term flares superposed to a variable component on longer timescales. The first flare is almost coincident with the $\gamma$-ray flare of 15 April 2011 (MJD 55667). As can be seen from the light curves, the optical behavior is similar to that observed at radio wavelengths shifted by approximately five months, with the optical emission occurring first. In both light curves, the period of high activity seems to have the same duration, taking into account the sampling limitation. We show this behavior in detail in Fig. \ref{7R}, where the optical light curve is shifted by 150 days and superposed to the radio data. This delay is confirmed by the discrete correlation function (DCF) analysis \citep{ede88}, presented in Fig. \ref{DCF}. In the same figure, we also present the DCF between 7mm and $\gamma$ rays and $R$ band and $\gamma$ rays, where no correlation was found  above the  3 $\sigma$ level.



\begin{figure*}
  \includegraphics[width=18cm]{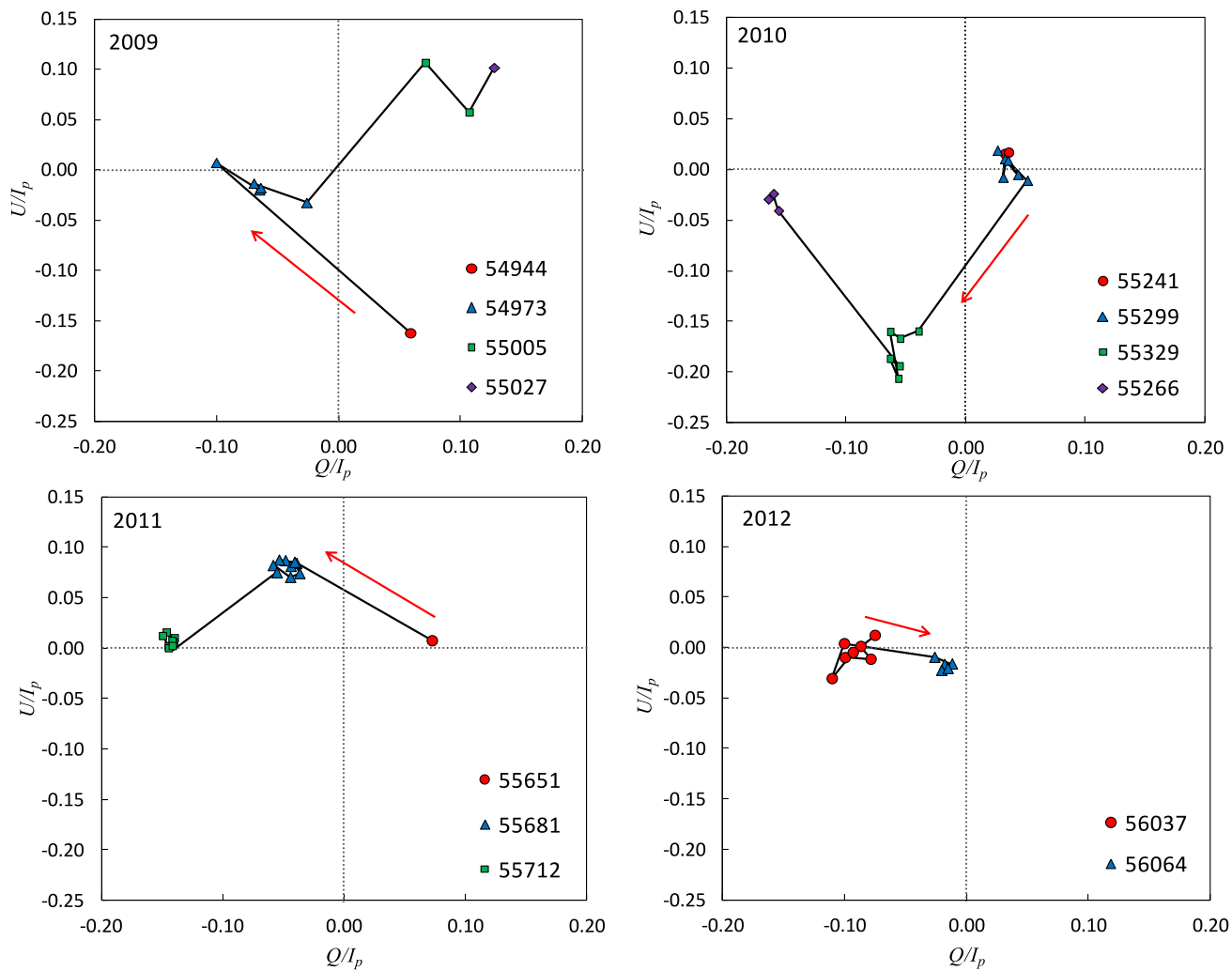}
  \caption{ $Q$ x $U$ Stokes plane normalized by $I_{p}$ during the four  years of our optical polarimetric monitoring. The arrows indicate the direction of increasing time.}
  \label{QU}
\end{figure*}

 \begin{figure}
\center
  \includegraphics[width=\columnwidth]{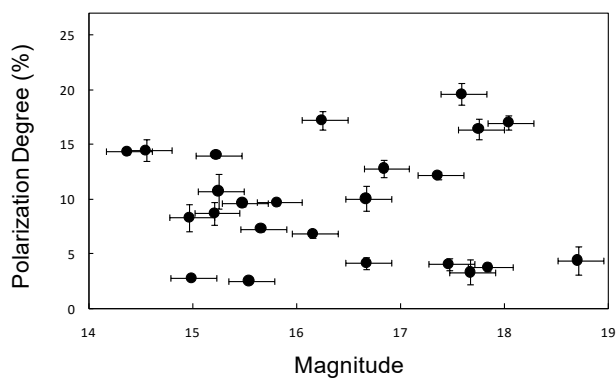}
  \caption{$PD$ vs. $R$-band magnitude recovered from our polarimetric images. No significant correlation was found based on the data.}
  \label{polmag}
\end{figure}

In the polarimetric data, at least for 2011, we see an increase in $PD$ and a rotation in $PA$ simultaneously  with the occurrence of the $\gamma$-ray flare of 15 April (MJD 55667).  On longer timescales, we verify the existence of a gradual $PA$ rotation in the $Q/I_p \times U/I_p$ plane,  as shown in  Fig. \ref{QU}. Because of the $\sin 2PA$ and $\cos 2PA$ dependence of the Stokes parameters, a continuous and gradual rotation in $PA$ will appear as  consecutive quadrant changes in that plane. In 2010, we detected a slow and gradual rotation of about $100^{\rm o}$ over six months and, although there are gaps that prevent an unambiguous determination of the direction of rotation, this behavior was also observed by \citet{ale14} and by \citet{kie16}. 



The relation between $PD$ and  total $R$ magnitude, obtained from our polarimetric observations, as described in section \ref{pol}, is shown in Fig. \ref{polmag}. We do not see any clear evidence of correlation, which was confirmed by a weak correlation coefficient of $r = 0.01$. Using data obtained between December 2009 and January 2010 (MJD 55173 to MJD 55206), \citet{ike11} found $r = 0.41^{+0.21}_{-0.27}$, using magnitudes in the $V$ band. However, based on observations during 2013 and 2014 where  a sequence of strong optical flares was observed with counterparts in $\gamma$-rays, \citet{ran18} found evidence of  an anti-correlation between optical polarization and flux density at $R$ band, with $r = 0.76$.


\section{Discussion}
\label{dis}

\subsection{Multiwavelength flux density variability}

The 7 mm (43 GHz) light curve of 3C~279 obtained with the Itapetinga radiotelescope, extending from the end of 2009 to March 2014  (MJD 55179 to MJD 56733), showed a very strong increase in activity at the end of 2011:   the flux density doubled over a period of two months, and remained at this level during almost one year. Although this blazar was monitored by several observatories, no report of this behavior was found in the literature, probably because it corresponded to a period of very low optical and $\gamma$-ray activity.

During 2010 (MJD 54650 to MJD 55400), a  good agreement was found between the Itapetinga light curve, the 43 GHz observations from Noto, and the 37 GHz observations from Mets\"ahovi, as reported by \citet{hay12}.
Also, the Itapetinga 43 GHz observations between February and July 2011  (MJD 55600 and MJD 55740) also match very closely those at 37 GHz from Mets\"ahovi Observatory, reported by \citet{ale14}.

A correlation was found for the whole period between our 7 mm light curve  and the $R$-band flux density from SMARTS, with a delay in radio of $170 \pm 30$ days relative to $R$ band, as expected from a compact, optically thick source that becomes optically thin as it expands \citep{mar85,tur99,tur00,bot88,bea14,bea17}.
A similar result was found by \citet{cha08} for the period 1996-2007,  using the $R$ band  and the 7 mm light curve of the core derived from  VLBA images,  obtaining  a delay of $130^{+30}_{-45}$ days. The larger delay in the single-dish data, although within the uncertainty interval, can be explained if the emitting source left the core before reaching its maximum flux density. This is certainly the case of the strong flares at the end of 2011, detected in our 7 mm  single-dish data,  as can be seen in Fig.\ref{all}. \citet{jor17}, in their analysis of the 43 GHz VLBA images for these epochs, found two components with intensities comparable to that of the core and very close to it. The position angles in the plane of the sky of these two components   were very different from those of the other jet components.

The formation epoch of the superluminal components can be determined from their kinematic properties. From the work of \citet{jor17} we were able to follow the evolution of one of these strong components, with PA $ 155^\circ \pm 5^\circ$ in the southeast direction, for which we found a  velocity of $6.2c$ and a maximum flux density of around 9 Jy. It was  formed around  MJD 55430 (2010 August 21), coinciding with the beginning of a $\gamma$-ray flare that reached maximum flux density in MJD 55365. This component was also observed by \citet{lu13} at 230 GHz with the Event Horizon Telescope (EHT) on 1-2 April 2011, when it was at a distance of 0.13 mas from the core. From MJD $\sim$ 55992 to MJD $\sim$ 56075, a stationary component was seen in the VLBA images from \citet{jor17}, at about a distance of 0.26~mas from the core, with position angle $175^\circ$  and mean flux density of about 18~Jy.  This component started moving at velocity $11c$ and  PA rotating from $-179^\circ$ to $-158^\circ$, while the flux density decreased from 20~Jy to 11~Jy in 6 months.

From the $\gamma$-ray light curve we can see, beginning in April 2011, a group of three flares (MJD 55667, 55697, and 55739) about 170 days before the beginning of the strong rise in the 7 mm light curve, which can be interpreted as  the superposition of the radio  counterparts of these $\gamma$-ray flares. The radio activity following this  group of three flares was closely correlated with strong optical flares, taking into account the corresponding delay and lasting for about a year, but does not seem to have any strong $\gamma$-ray counterparts.

No correlation was found in the DCF between radio--$\gamma$-ray and $R$-band--$\gamma$-ray light curves, although this result does not mean that the emission  is not produced in the same region. If the high energy emission is due to EC emission, it can be absorbed by the photons of the broad line region (BLR) if emitted very close to the core \citep{bott16}, or it can be missing due to the lack of low energy photons in the beam direction.

 \begin{figure}
  \includegraphics[width=\columnwidth]{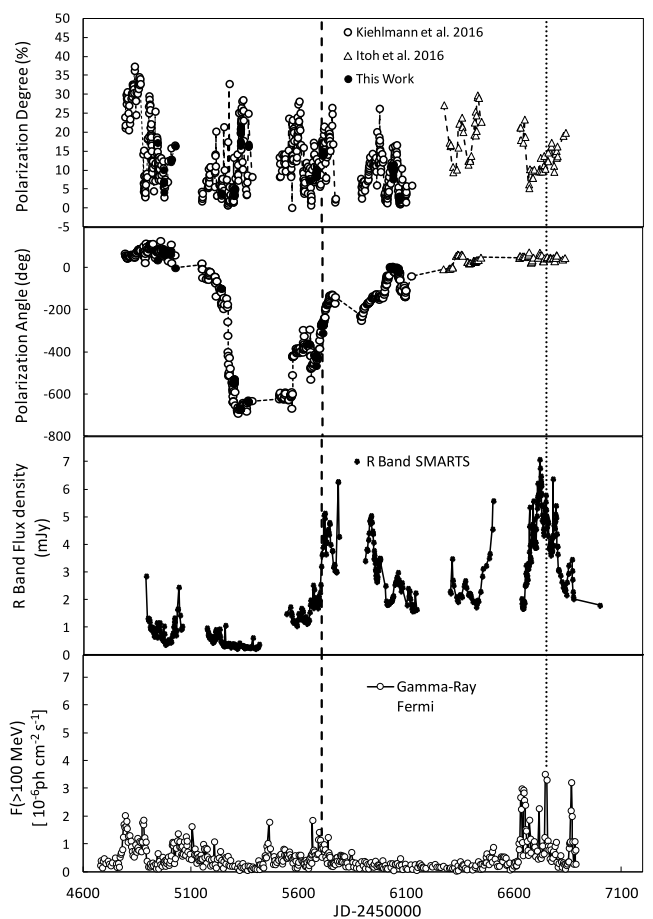} 
  \caption{Top: $PA$ variability obtained by \citet{kie16} adding the data of this work. As a criterion to solve the $180^{\rm o}$ multiplicity we minimized the difference between successive data points. Bottom: $R$-band and $\gamma$-ray variability.The dashed line indicates the $\gamma$ ray associated with a new jet component reported by \citet{jor17} that produced a $PA$ rotation, while the dotted line indicates a new jet component seen in VLBI images \citep{ran18} that did not produce changes in $PA$.}
  \label{PAall}
\end{figure}

\subsection{Jet precession}

The large variation $(\sim 80^\circ)$ in the PA of the superluminal components formed during 2010-2012, together with the difference in their superluminal velocities, seems to support the idea of jet precession, as suggested by \citet{abr98} and \citet{qui11}.
The epoch in which the components are brighter and the variation in the position angles is larger and faster must correspond to the epoch when the angle between the jet and the line of sight is the smallest, as seems to have occurred between 2010 and 2012. Considering the precession model suggested by \citet{abr98}, with a period of 25 years, half the periodicity found by \citet{web90}, the epoch of closest approach would correspond to the year 2012. However \citet{qui11} also found another large variation in the position angle of the superluminal components around 2006, which is compatible with a precession period of only 6 years, which is close to the period in the infrared found by \citet{fan99}. A longer and detailed analysis of the kinematics of the superluminal components would be necessary to confirm and fully understand jet precession.

\subsection{Polarimetric variability}

The existence of several components contributing to the total emission and their evolution along the jet can also present signatures on the polarimetric variability. A new polarized component can produce large changes in polarization parameters, but small changes are also possible depending on the values of the Stokes parameters of the jet and the new component \citep{bea17}. In 2011, we detected in our polarimetric data a gradual rotation of almost $200^{\rm o}$ in $PA$ at the epoch of highest activity in the optical light curve. Around the same epoch, \citet{ale14} found a rotation of about $140^{\rm o}$, while \citet{kie16} obtained $352^{\rm o}$. These differences can be explained by  gaps in the polarimetric monitoring, which  can produce spurious rotation values \citep{kie13,kie16,bea17,bea18}.



In order to improve the data coverage of our analysis, we considered our data together with the $PA$ measurements from \citet{kie16} and  from \citet{ito16}. In the upper part of Fig. \ref{PAall} we show the whole dataset that was used to solve the $180^{\rm o}$ multiplicity with the simplest assumption that between two consecutive observations $PA$ changed by the lowest possible value, allowing both clockwise and counterclockwise rotation. This is an acceptable interpretation in the absence of large time lags between observations \citep{bea18}. We present the results in the second panel of Fig. \ref{PAall}, where it is possible to see that the $PA$ rotated by almost $800^{\rm o}$  clockwise  between MJD 54100 and 55300, without any clear counterpart in variability of the total flux density at $R$ band and $\gamma$ rays. After that large rotation, the $PA$ rotated again by almost the same quantity and in a similar time interval, but counterclockwise, until it reached a level similar to that at the beginning of the monitoring.  




This behavior of two directions of $PA$ rotation was already detected in PKS~1510-089 \citep{bea17} and OJ~287 \citep{coh18}. In 3C~279,  $PA$ rotations were interpreted by \citet{ale14} as a consequence of the bending of the jet. We suggest that the rotations observed between 2010 and 2011 are  the consequence of the ejection of a series of jet components during that epoch, as observed in the VLBI images reported by \citet{jor17} and identified by us in Section 4.1 . 




\subsection{Polarimetric variability caused by the ejection of a new jet component}

We  followed the two component scenario described by \citet{hol84}  and applied it to describe the polarization variability due to the appearance of a new jet component. According to what was done in \citet{bea17} in the case of PKS~1510-089, we estimated the possible values of $PD$ and $PA$  of a new component that produced the observed change in  $PA$. 

To this end, we attributed a set of  initial Stokes parameters for the whole jet, considering the emission of all the individual components. We  computed the Stokes parameters after the emission of a new component using the following equations:

\begin{equation}
\label{eq3}
Q_{final} = Q_{jet}+Q_{new}
,\end{equation}
\label{eq4}
\begin{equation}
U_{final} = U_{jet}+U_{new}
,\end{equation}
\begin{equation}
\label{eq5}
I_{p,final} = \sqrt{Q_{final}^{2}+U_{final}^{2}}
,\end{equation}
where $Q$, $U$, and $I_p$ are the Stokes parameters and the indexes $jet$, $final$, and $new$ correspond to the quantities before and after the formation of a new component, and those of the new component, respectively.  In principle, the polarimetric parameters $U$ and $Q$ of the new component can differ from the previous ones due to differences in magnetic field, particle densities, and/or geometry. We can write each parameter as a function of $PA$:
\begin{equation}
\label{eq1}
Q_{jet} = I_{p,jet}\cos{2 \theta_{jet}},
\end{equation}
\begin{equation}
\label{eq2}
U_{jet} = I_{p,jet}\sin{2 \theta_{jet}},
\end{equation}
\noindent
where $\theta$ is the $PA$ and the indexes remain the same. Solving the equation we obtain:
\begin{equation}
\label{eq6}
\cos{2 \theta_{final}} = (I_{p,jet}\cos{2 \theta_{jet}}+I_{p,new}\cos{2 \theta_{new}})/I_{p,final}
,\end{equation}
\begin{equation}
\label{eq7}
\sin{2 \theta_{final}} = (I_{p,jet}\sin{2 \theta_{jet}}+I_{p,new}\sin{2 \theta_{new}})/I_{p,final}
,\end{equation}
\begin{equation}
\label{eq8}
I_{p,final}^{2} = I_{p,jet}^{2}+I_{p,new}^{2}+2I_{p,jet}I_{p,new} \cos{2(\theta_{new}-\theta_{jet})}   
.\end{equation}

Considering the epoch between 1 and 30  May 2011  (MJD 55693 and MJD 55711), the  $PA$ rotated from $117\fdg0$ to $92\fdg4$, while the  $PD$ increased from $9.3\%$ to $14\%$. This variation could be attributed to a new component with a polarized optical flux of $(0.028 \pm 0.019)$ mJy and a $PA$ of $138\fdg0 \pm 2\fdg2$.  This new component appeared in the  VLBI images presented by \citet{jor17} as discussed in Sec. 4.1.  The ejection of a new component with such characteristics could change the direction of rotation, as shown by the dashed line in Fig. \ref{PAall}. On the other hand, as shown by \citet{bea17}, a new component might not   produce  changes in $PA$, as is the case of NC2 \citep{ran18}  associated with the $\gamma$-ray flare in  MJD 56756, indicated by the dotted line in Fig. \ref{PAall}.

\section{Conclusions}
\label{conclusion}

We report five years of observations of 3C~279 at the 7 mm radio continuum and  $R$-band polarimetry, obtained at the Itapetinga and Pico dos Dias observatories, respectively. We compared our monitoring with $R$-band total flux density obtained from the SMARTS program and $\gamma$-ray data from Fermi/LAT. We also used data from VLBA-BU-Blazar at 7 mm to compare our single-dish results with interferometric images, and we analyzed polarimetric variability together with the results reported by \citet{kie16}. We can summarize our conclusions as follows:

\begin{itemize}
\item The 7~mm observations revealed a period of very high flux density starting at the end of 2011 and lasting for about a year.
\item The 7~mm and $R$-band light curve from SMARTS showed very good correlation, with a lag of  of $170 \pm 30$ days between radio and optical data.
\item This delay also allowed us  to associate a group of $\gamma$-ray flares  observed in April 2011 with the rise, by a factor of two, in the 7~mm flux density at the end of 2011. However, no correlation was found between the 7~mm and $\gamma$-ray light curves for the whole period.
\item During the occurrence of these $\gamma$-ray flares, polarimetric data showed an increase in $PD$ and a large rotation in $PA$, which was interpreted as a consequence of the superposition of the polarimetric parameters of the jet and of a new component. The computed  polarimetric parameters of this new component are $(0.028 \pm 0.019)$ mJy for the polarized flux and   $138\fdg0 \pm 2\fdg2$ for $PA$.
\item VLBA images showed, at the epoch of the 7~mm maximum, the appearance of two new components in the jet, ejected from the core with velocities of $6c$ and $11c$, with PAs in the plane of the sky very different from those of the other jet components.
\item The differences in velocities and PAs of the different components are compatible with  the jet precession models of \citet{abr98} and \citet{qui11}, with a period of 25 years in the observer reference frame and  maximum approach to the line of sight occurring in 2011. 

\end{itemize}


\begin{acknowledgements}
We are grateful to the Brazilian research agencies FAPESP and CNPq for financial support (FAPESP Projects: 2008/11382-3 and 2014/07460-0). We thank INPE (Instituto Nacional de Pesquisas Espaciais) for the operation of ROI (Radio Observat\'{o}rio do Itapetinga).
This study makes use of 43 GHz VLBA data from the VLBA-BU Blazar Monitoring Program (VLBA-BU-BLAZAR; http://www.bu.edu/blazars/VLBAproject.html), funded by NASA through the Fermi Guest Investigator Program. The VLBA is an instrument of the National Radio Astronomy Observatory. The National Radio Astronomy Observatory is a facility of the National Science Foundation operated by Associated Universities, Inc. This paper has made use of up-to-date SMARTS optical/near-infrared light curves that are available at www.astro.yale.edu/smarts/glast/home.php.

\end{acknowledgements}

%
%

\longtab{

\begin{longtable}{cccccccccccc}

    \caption{ {\it R}-flux density and polarization of 3C~279 obtained in this work \label{151}}\\
         \hline\hline

        Date & JD & Q/I & U/I   & Total Flux & Error & Polarized Flux & Error &  \it{PD} & Error & \it {PA}  & Error \\
     & -2450000 & & &  mJy & mJy  & mJy & mJy & $ \% $ & $ \% $ & degrees & degrees \\
    \hline
    \endfirsthead           
        \caption{continued.}\\
        \hline\hline
    \endhead

2009-04-22  &  4944.56  &  0.059  &  -0.162  &  0.94  &  0.28  &  0.16  &  0.04  &  17.2  &  0.8  &  35.2  &  1.4  \\
  &    &    &    &    &    &    &    &    &    &    &    \\
2009-05-20  &  4972.51  &  -0.100  &  0.007  &  0.63  &  0.19  &  0.06  &  0.03  &  10.1  &  1.1  &  87.6  &  3.2  \\
2009-05-21  &  4973.47  &  -0.065  &  -0.020  &  1.01  &  0.30  &  0.07  &  0.02  &  6.8  &  0.2  &  76.0  &  0.9  \\
2009-05-21  &  4973.53  &  -0.064  &  -0.018  &  1.01  &  0.30  &  0.07  &  0.02  &  6.6  &  0.3  &  76.8  &  1.2  \\
2009-05-21  &  4973.59  &  -0.070  &  -0.013  &  1.01  &  0.30  &  0.07  &  0.02  &  7.1  &  0.5  &  79.1  &  2.0  \\
2009-05-22  &  4974.50  &  -0.026  &  -0.032  &  0.63  &  0.19  &  0.03  &  0.01  &  4.2  &  0.5  &  58.1  &  3.7  \\
  &    &    &    &    &    &    &    &    &    &    &    \\
2009-06-22  &  5005.45  &  0.071  &  0.107  &  0.54  &  0.16  &  0.07  &  0.03  &  12.8  &  0.8  &  63.0  &  1.8  \\
2009-06-23  &  5006.44  &  0.107  &  0.058  &  -  &  -  &  -  &  -  &  12.2  &  0.3  &  80.3  &  0.8  \\
  &    &    &    &    &    &    &    &    &    &    &    \\
2009-07-14  &  5027.43  &  0.127  &  0.102  &  0.33  &  0.10  &  0.05  &  0.04  &  16.3  &  1.1  &  178.4  &  1.9  \\
  &    &    &    &    &    &    &    &    &    &    &    \\
2010-02-13  &  5241.78  &  0.032  &  0.016  &  0.21  &  0.06  &  0.01  &  0.01  &  3.6  &  0.4  &  80.5  &  3.5  \\
2010-02-13  &  5241.81  &  0.036  &  0.017  &  0.21  &  0.06  &  0.01  &  0.01  &  4.0  &  0.3  &  80.7  &  2.3  \\
  &    &    &    &    &    &    &    &    &    &    &    \\
2010-04-12  &  5299.60  &  0.031  &  -0.008  &  0.25  &  0.08  &  0.01  &  0.01  &  3.2  &  1.0  &  190.9  &  8.6  \\
2010-04-12  &  5299.64  &  0.032  &  0.011  &  0.25  &  0.08  &  0.01  &  0.01  &  3.4  &  1.3  &  174.9  &  11.2  \\
2010-04-13  &  5300.62  &  0.044  &  -0.005  &  0.10  &  0.03  &  0.00  &  0.01  &  4.4  &  1.3  &  189.8  &  8.4  \\
2010-04-14  &  5301.57  &  0.026  &  0.019  &  0.30  &  0.09  &  0.01  &  0.01  &  3.2  &  0.1  &  168.8  &  0.9  \\
2010-04-14  &  5301.59  &  0.035  &  0.009  &  0.30  &  0.09  &  0.01  &  0.01  &  3.6  &  0.7  &  179.4  &  5.4  \\
2010-04-14  &  5301.62  &  0.051  &  -0.011  &  0.30  &  0.09  &  0.02  &  0.01  &  5.3  &  0.9  &  12.5  &  4.7  \\
  &    &    &    &    &    &    &    &    &    &    &    \\
2010-05-12  &  5329.52  &  -0.040  &  -0.160  &  0.18  &  0.05  &  0.03  &  0.03  &  16.5  &  0.9  &  46.3  &  1.5  \\
2010-05-12  &  5329.62  &  -0.055  &  -0.167  &  0.18  &  0.05  &  0.03  &  0.04  &  17.6  &  0.4  &  48.4  &  0.6  \\
2010-05-13  &  5330.52  &  -0.063  &  -0.160  &  0.27  &  0.08  &  0.05  &  0.04  &  17.2  &  0.9  &  50.7  &  1.4  \\
2010-05-13  &  5330.54  &  -0.056  &  -0.206  &  0.27  &  0.08  &  0.06  &  0.05  &  21.4  &  1.3  &  47.6  &  1.7  \\
2010-05-13  &  5330.59  &  -0.063  &  -0.187  &  0.27  &  0.08  &  0.05  &  0.04  &  19.7  &  0.8  &  49.3  &  1.1  \\
2010-05-13  &  5330.62  &  -0.055  &  -0.194  &  0.27  &  0.08  &  0.05  &  0.04  &  20.2  &  1.0  &  47.9  &  1.4  \\
  &    &    &    &    &    &    &    &    &    &    &    \\
2010-06-18  &  5366.44  &  -0.157  &  -0.041  &  0.23  &  0.07  &  0.04  &  0.03  &  16.2  &  0.8  &  82.0  &  1.4  \\
2010-06-18  &  5366.47  &  -0.161  &  -0.024  &  0.23  &  0.07  &  0.04  &  0.04  &  16.3  &  1.2  &  85.0  &  2.2  \\
2010-06-18  &  5366.51  &  -0.165  &  -0.029  &  0.23  &  0.07  &  0.04  &  0.04  &  16.8  &  0.9  &  84.2  &  1.5  \\
  &    &    &    &    &    &    &    &    &    &    &    \\
2011-03-30  &  5651.63  &  0.073  &  0.008  &  1.61  &  0.48  &  0.12  &  0.02  &  7.3  &  0.2  &  175.7  &  0.9  \\
  &    &    &    &    &    &    &    &    &    &    &    \\
2011-04-29  &  5681.47  &  -0.039  &  0.084  &  1.89  &  0.57  &  0.18  &  0.02  &  9.3  &  0.0  &  123.3  &  0.0  \\
2011-04-29  &  5681.50  &  -0.048  &  0.087  &  1.89  &  0.57  &  0.19  &  0.02  &  9.9  &  0.2  &  121.2  &  0.6  \\
2011-04-29  &  5681.53  &  -0.053  &  0.088  &  1.89  &  0.57  &  0.19  &  0.04  &  10.3  &  0.8  &  120.0  &  2.3  \\
2011-04-29  &  5681.56  &  -0.044  &  0.081  &  1.89  &  0.57  &  0.17  &  0.02  &  9.2  &  0.2  &  121.6  &  0.6  \\
2011-04-29  &  5681.58  &  -0.041  &  0.085  &  1.89  &  0.57  &  0.18  &  0.02  &  9.5  &  0.3  &  123.0  &  0.9  \\
2011-04-30  &  5682.53  &  -0.037  &  0.074  &  3.03  &  0.91  &  0.25  &  0.07  &  8.3  &  1.7  &  122.2  &  5.9  \\
2011-04-30  &  5682.57  &  -0.044  &  0.070  &  3.03  &  0.91  &  0.25  &  0.05  &  8.3  &  1.2  &  119.3  &  4.1  \\
      2011-05-01  &  5683.57  &  -0.058  &  0.082  &  1.39  &  0.42  &  0.14  &  0.02  &  10.1  &  0.3  &  117.5  &  1.0  \\
2011-05-01  &  5683.60  &  -0.055  &  0.075  &  1.39  &  0.42  &  0.13  &  0.02  &  9.3  &  0.1  &  117.0  &  0.4  \\
  &    &    &    &    &    &    &    &    &    &    &    \\
2011-05-30  &  5712.47  &  -0.140  &  0.010  &  2.39  &  0.72  &  0.33  &  0.04  &  14.0  &  0.3  &  92.4  &  0.7  \\
2011-05-30  &  5712.50  &  -0.140  &  0.005  &  2.39  &  0.72  &  0.34  &  0.03  &  14.0  &  0.2  &  91.5  &  0.4  \\
2011-05-31  &  5713.43  &  -0.142  &  0.008  &  4.46  &  1.34  &  0.63  &  0.13  &  14.2  &  2.3  &  91.1  &  4.6  \\
2011-05-31  &  5713.48  &  -0.146  &  0.016  &  4.46  &  1.34  &  0.66  &  0.03  &  14.7  &  0.1  &  92.7  &  0.2  \\
2011-05-31  &  5713.54  &  -0.149  &  0.012  &  4.46  &  1.34  &  0.67  &  0.04  &  15.0  &  0.1  &  92.0  &  0.2  \\  
2011-06-01  &  5714.43  &  -0.145  &  0.002  &  5.28  &  1.58  &  0.76  &  0.05  &  14.5  &  0.3  &  89.5  &  0.7  \\
2011-06-01  &  5714.46  &  -0.144  &  0.000  &  5.28  &  1.58  &  0.76  &  0.03  &  14.4  &  0.1  &  89.1  &  0.1  \\
2011-06-01  &  5714.49  &  -0.141  &  0.002  &  5.28  &  1.58  &  0.75  &  0.05  &  14.1  &  0.3  &  89.5  &  0.7  \\
  &    &    &    &    &    &    &    &    &    &    &    \\
2012-04-19  &  6037.50  &  -0.075  &  0.012  &  2.42  &  0.73  &  0.18  &  0.08  &  7.6  &  2.7  &  186.5  &  10.0  \\
2012-04-19  &  6037.55  &  -0.087  &  0.001  &  2.42  &  0.73  &  0.21  &  0.04  &  8.7  &  1.0  &  182.3  &  3.4  \\
2012-04-19  &  6037.56  &  -0.078  &  -0.012  &  2.42  &  0.73  &  0.19  &  0.04  &  7.9  &  1.1  &  177.7  &  4.1  \\
2012-04-19  &  6037.60  &  -0.100  &  -0.010  &  2.42  &  0.73  &  0.24  &  0.03  &  10.0  &  0.3  &  179.1  &  0.8  \\
2012-04-19  &  6037.63  &  -0.093  &  -0.005  &  2.42  &  0.73  &  0.23  &  0.02  &  9.3  &  0.2  &  0.4  &  0.7  \\
2012-04-22  &  6040.48  &  -0.110  &  -0.031  &  2.34  &  0.70  &  0.27  &  0.02  &  11.5  &  0.0  &  172.1  &  0.1  \\
2012-04-22  &  6040.53  &  -0.100  &  0.004  &  2.34  &  0.70  &  0.23  &  0.09  &  10.0  &  3.2  &  181.0  &  9.1  \\
  &    &    &    &    &    &    &    &    &    &    &    \\
2012-05-16  &  6064.46  &  -0.026  &  -0.010  &  2.98  &  0.89  &  0.08  &  0.01  &  2.8  &  0.2  &  177.9  &  1.6  \\
2012-05-17  &  6065.46  &  -0.012  &  -0.016  &  1.79  &  0.54  &  0.04  &  0.01  &  2.0  &  0.2  &  159.9  &  2.3  \\
2012-05-17  &  6065.49  &  -0.018  &  -0.017  &  1.79  &  0.54  &  0.04  &  0.01  &  2.5  &  0.2  &  165.5  &  2.4  \\
2012-05-17  &  6065.52  &  -0.020  &  -0.023  &  1.79  &  0.54  &  0.05  &  0.01  &  3.1  &  0.1  &  163.1  &  1.0  \\
2012-05-17  &  6065.55  &  -0.015  &  -0.021  &  1.79  &  0.54  &  0.05  &  0.01  &  2.6  &  0.1  &  160.2  &  1.2  \\

\hline

\hfill{}

\end{longtable}
\tablefoot{{\it PA} values have a multiplicity of $+n180^\circ$.}

}

\end{document}